\documentstyle[twocolumn,prb,aps]{revtex}

\begin{document}
\draft
\preprint{Dortmund, August 1995}

%
\title{Control of the finite size corrections
       in exact diagonalization studies}
\author{Claudius Gros
       }
\address{Institut f\"ur Physik, Universit\"at Dortmund,
         44221 Dortmund, Germany
        }
\date{\today}
\maketitle
\begin{abstract}
The finite size corrections in exact diagonalization
studies of the one- and two dimensional Hubbard model can
be reduced systematically by a grand-canonical integration
over boundary conditions.
We present results for ground-state properties
of the 2D Hubbard model and an evaluation of the
specific heat for the 1D and 2D Hubbard model.
We find the reduction of the finite size
corrections to be substantial, especially in 2D.
\end{abstract}
%

\section*{Introduction}

The exact diagonalization of the Hubbard model
(or the $t-J$ model) on finite clusters has
become one of the prominent techniques in the
study of correlated electron systems \cite{dagotto_94}.
This technique has the advantage of
treating the correlations on finite clusters
without any approximation. Since the Hilbert space
grows exponentially with cluster size
one is limited, for the Hubbard model,
to clusters with up $N_s=10$ sites. By
{\sl finite size corrections} one means
the difference between the
results obtained for the finite
cluster and its (unkown) value in the
thermodynamic limit, $N_s=\infty$.

The control of the finite size corrections poses
a great challenge to the exact diagonalization technique,
especially in the case of two dimensions (2D).
In 1D, it is a standard procedure to plot the results
systematically as a function of chain length,
$N_s = 6, 8, 10, 12, \dots\ $ sites.
In 2D, on the other side, where most calculations have been done
up to now with either periodic or antiperiodic boundary conditions,
it is difficult to estimate the finite size corrections by direct
comparison of clusters of different sizes, for
two reasons:

\begin{enumerate}
\item The finite size corrections are often non-monotonic
      as a function of $N_s$, due to strongly varying
      cluster geometries.

\item The nominal density of particles, $n=N_e/N_s$
      does normally not coincide for clusters of different
      sizes as the allowed number of particles
      $N_e=1,2,3,\dots$ is an integer.
\end{enumerate}

Both of these two difficulties can be circumvented
by a method introduced recently \cite{gros_92},
the {\it integration over boundary conditions} (IBC).
The IBC circumvents the first of above difficulties
by performing a grand-canonical integration over boundary
conditions.
It has been shown, that this procedure removes all those
finite size corrections which are caused by the special
geometry of the Fermi-sea of the cluster \cite{note_1}.
The {\it grand-canonical} approach then leads to the
possibility of directly comparing the results for different
$N_s$ with the {\it same} density $n$, solving also
the second of above difficulties.

In a previous publication \cite{gros_92} the
IBC had been introduced and some results for the
Hubbard model on very small clusters had been
presented. Here we want to study in detail how
the finite size corrections can be estimated
{\it systematically} within the IBC both at zero
and at finite temperatures. For this purpose
we will concentrate on the Hubbard model,
the same study could be performed, in principle,
for the $t-J$ model or any other cluster Hamiltonian.


\section*{integration over boundary conditions}

We consider the Hubbard Hamiltonian
on a cluster with $N_s$ sites,
\begin{equation}
H = \sum_{<i,j>,\sigma}
 \left( t_{i,j}\,
        c_{i,\sigma}^\dagger c_{j,\sigma}^{\phantom{\dagger}}
      + t_{i,j}^*\,
        c_{j,\sigma}^\dagger c_{i,\sigma}^{\phantom{\dagger}} \right)
  + U\sum_i n_{i,\uparrow}n_{i,\downarrow}
\label{H}
\end{equation}
where the symbol $<i,j>$ denotes pairs of n.n. sites and where
the  $c_{i,\sigma}^\dagger$ and the $c_{i,\sigma}^{\phantom{\dagger}}$
are the electron creation/destruction operators of
spin $\sigma = \uparrow,\downarrow$ respectively.
The hopping amplitudes,
\begin{equation}
  t_{i,j} \equiv -t\, \exp[i\alpha_{i,j}]
\label{t}
\end{equation}
depend, in general, on phases $\alpha_{i,j}$, related
to the boundary condition. In the following all energies
will be measured in unities of $t$.
We will now give an account of the IBC for 1D, the generalization
to 2D is then straightforward.

In 1D we can choose the $\alpha_{i,j}=\alpha/N_s$, where
$\alpha\in[0,2\pi[$ is the boundary condition.
Periodic and antiperiodic boundary conditions
correspond to $\alpha=0$ and $\alpha=\pi$ respectively.
The IBC technique needs the exact diagonalization of the cluster
for any particle number $N_e=0,1,2,\dots,2N_s$ and any
$\alpha\in[0,2\pi[$. For any given $\alpha$ one then
calculates the free energy
\begin{equation}
  F_\alpha(\mu) =- T k_B\ln(\sum_{N_e=0}^{2N_s}
       e^{\beta\mu N_e}\sum_{k=0}^{N_H-1} e^{-\beta E_k(\alpha,N_e)}),
\label{F_alpha}
\end{equation}
where $\beta=1/(k_B T)$ is the inverse temperature, $\mu$ the
chemical potential, $N_H-1$ the dimension of the
Hilbert space and the $E_k(\alpha,N_e)$ the eigenenergies.
($N_H$ depends on both on $N_s$ and $N_e$).
The total free energy of the cluster and the particle
density $n\in[0,2]$ are then given by the average
\begin{equation}
  F(\mu) = \int_0^{2\pi} {d\alpha\over 2\pi} F_\alpha(\mu).
\label{F_total}
\end{equation}
and by $n=-{1\over N_s}{\partial F(\mu)\over\partial\mu}$.
The free energy Eq.\ (\ref{F_total}) obtained with
the IBC technique is constructed in such a way that the finite
size corrections {\it vanish identically} in the limit
$U=0$ \cite{gros_92}. One can understand this result most
easily in momentum space, noting that the Fermi-sea, as
obtained by the IBC,
has no finite size corrections for $U=0$. This property
of the IBC holds in any dimension and is especially important
in 2D. The particle density $n$ can furthermore be tuned
to {\it any} value, by choosing an appropiate $\mu$,
allowing to directly compare data from
clusters with different $N_s$.
These properties distingish the IBC technique
from other exact diagonalization studies employing a range of
boundary conditions \cite{poilblanc_91}.

The integration over boundary conditions, appearing in
Eq.\ (\ref{F_total}), is replaced in actual calculations
by a finite sum over $N_\alpha$ equal-distant boundary conditions.
Indeed we will find further below that only a few
$N_\alpha$ are needed in order to already obtain
most of the reduction in the finite size corrections.
Ground-state properties can be calculated, as usual,
by the Lanczos technique. With $E_0(\alpha,N_e)$ being the
ground-state energy for a given $\alpha$ and $N_e$ one
finds the estimate $e_0(\mu)$ for the grand-canonical
ground-state energy per site to be
\begin{equation}
e_0(\mu) = {1\over N_s N_\alpha} \sum_{\alpha}\,
           \inf_{N_e}\, [E_0(\alpha,N_e)-\mu N_e].
\label{e_0}
\end{equation}
A Legendre transformation then yields the canonical ground-state
energy of the cluster, $e_0(n,N_s)$, as a function of density.
In Fig.\ \ref{2D_U8} we have
plotted results obtained for the 2D Hubbard model with $U=4t$.
The results for $N_\alpha=1$, corresponding to periodic boundary
conditions only, differ typically by
$10\%-20\%$ in between clusters with $N_s=8$ and $N_s=10$ sites.
Also shown in Fig.\ \ref{2D_U8} are the data obtained
for $N_\alpha=25$ \cite{note_2}. Here the data differ
only by $1\%-2\%$ in between $N_s=8$ and $N_s=10$, an
improvement by one order of magnitude. For comparision
we have plotted also data obtained by projection Monte
Carlo for a $10\times10$ system with periodic boundary condition
by Furukawa and Imada \cite{F_I}.


\section*{specific heat}

Eq.\ (\ref{F_alpha}) for the free energy
needs the knowledge of all eigenstates
and can therefore be used only for very small sytems,
typically up to $N_s=6$ for the Hubbard model.
Several approximative methods have been developed
for the numerical evaluation of thermodynamic properties of
sytems with large Hilbert spaces \cite{roeder_95,prelovsek_94}.
Here we use the {\sl kernel polynomial approximation} (KPA)
developed by Silver and Roeder \cite{roeder_95}, since it has a
very good error control. One starts by scaling the
Hamiltonian so that the magnitude of all eigenenergies
is less than unity. The
grand-canonical partition function,
$Z_\alpha(\mu) = \exp[-\beta F_\alpha(\mu)]$,
is then expressed as an integral over the density of states,
$D_\alpha(\omega,N_e)=
1/N_H\sum_{k=0}^{N_H-1}\delta(\omega-E_k(\alpha,N_e))$,
\begin{equation}
Z_\alpha(\mu) = \sum_{N_e=0}^{2N_s} N_H\,e^{\beta\mu}
\int_{-1}^{1} d\omega D_\alpha(\omega,N_e) e^{-\beta\omega}.
\label{Z}
\end{equation}
The density of states can be expanded in a set of
orthogonal polyomials, here we take Legendre
polyomials, $P_l(\omega)$:
\begin{equation}
D_\alpha(\omega,N_e)=
\sum_{l=0}^{N_{mom}}\ a_l(\alpha,N_e)\, P_l(\omega),
\label{D}
\end{equation}
with $N_{mom}\rightarrow\infty$. Using the orthogonality
relations for Legendre polynomials one can express the
coefficients $a_l(\alpha,N_e)$ in
term of the eigenstates of the Hamiltonian,
$H|E_k(\alpha,N_e)\rangle= E_k(\alpha,N_e)|E_k(\alpha,N_e)\rangle$,
via
\begin{equation}
a_l(\alpha,N_e)= {2l+1\over2N_H}\sum_{k=0}^{N_H-1}
\langle E_k(\alpha,N_e)|P_l(H)|
           E_k(\alpha,N_e)\rangle.
\label{a_l}
\end{equation}
In practice one uses a finite $N_{mom}<\infty$. A finite values
of $N_{mom}$ leads to the well known Gibbs oscillations in
$D_\alpha(\omega,N_e)$. The Gibbs oscillations can be smoothed
out by the replacement $a_l\rightarrow a_l\,g_l(N_{mom})$ in (\ref{D}).
The optimal functional form of the smoothing functions
$g_l(N_{mom})$, with $g_0(N_{mom})=1$ and $g_{N_{mom}+1}(N_{mom})=0$, has
been studied intensively in the literature \cite{roeder_95}.
We use $g_l(N_{mom}) = (sin(z)/z)^3$, with $z=l\pi/(N_{mom}+1)$.

The number $N_{mom}$ of polynomials necessary for an accurate
representation of the density of states in (\ref{D}) increases
with decreasing temperature, $T$. We have evaluated the
specific heat $c_V=\beta^2<(H-<H>)^2>$ for a 6-site
Hubbard chain with periodic boundary conditions
exactly, via Eq. (\ref{F_alpha}), and via the kernel
polynomial approximation with various $N_{mom}$.
We have plotted in Fig.\ \ref{1D6} the results for
(a) $U=2$ and (b) $U=16$. F
The KPA becomes asymptotically exact for large temperatures
and any values of $N_{mom}$. A numerical accurate
approximation to the specific heat for temperatures
down to $T\sim0.25t$ may be obtained with
$N_{mom}\sim10^3$ and $N_{mom}\sim10^4$, for
$U=2$ and $U=16$ respectively.
The large value, $N_{mom}\sim10^4$, needed for big
$U$'s is the reason that all data presented
further below will be for $U=2$.

Formula (\ref{a_l}) is clearly not useful for larger
clusters. It has been observed \cite{roeder_95} that one
can {\sl systematically} approximate the trace occuring
in the RHS side of (\ref{a_l}) by sampling over
$N_r$ random states $|r\rangle$,
\begin{equation}
a_l(\alpha,N_e)\approx {2l+1\over2N_r}\sum_{r=1}^{N_r}
\langle r|P_l(H)|r\rangle.
\label{a_random}
\end{equation}
It is possible to evaluate (\ref{a_random})
recursively, by a procedure very similar to the
Lanczos technique \cite{roeder_95}.
The errors introduced by random averaging
vanish like $1/\sqrt{N_r}$ \cite{roeder_95}.
We have performed extensive tests of the dependence
of the data on $N_r$; we will present further
below only data which are unaffected by finite-$N_r$
effects.


\section*{results}

In Fig.\ \ref{1D_U2} we have plotted the results for the
specific heat per site for various 1D chains of length
$N_s=4,6,8,10$, both for the case of periodic boundary
conditions only, $N_\alpha=1$, and for $N_\alpha=10$.
Let us first note that the specific-heat curves for
clusters of different sizes have eventually all to coincide
at large enough temperatures, due to the grand-canonical
formulation. At large enough temperatures the leading terms
contributing to the specific heat of a cluster of a given
size are identical to the leading terms of a high-temperature
expansions. One can consequently define a $T^*(N_s,\epsilon)$ as the
temperature above which the finite size corrections to the
specific heat are smaller than a given accuracy $\epsilon$,
say $\epsilon\sim1\%$. As the specific heat for $N_s=\infty$
is generally not known we took as a practical estimate for
$T^*(N_s,\epsilon)$ the criterium that
$|c_v(T,N_s)-c_v(T,N_s-2)|<\epsilon$ for all $T>T^*(N_s,\epsilon)$.
We found that $T^*(N_s,\epsilon)$ is roughly inversely proportional
to $N_s$, for the 1D data presented in Fig.\ \ref{1D_U2} and that
the integration over boundary conditions reduces
$T^*(N_s,\epsilon)$ by factors of $2-3$, as it is evident from
a comparison of Fig.\ \ref{1D_U2}(a) and Fig.\ \ref{1D_U2}(b).

The IBC improves the cluster estimates of the specific heat in
a second way, besides the reduction of $T^*(N_s,\epsilon)$.
One finds empirically that the specific heat becomes {\sl linear}
at low temperatures in the limit $N_\alpha\rightarrow\infty$.
This property of the low-T specific heat is present for
all $N_s$ and is a consequence of the absence of those
finite size corrections which are caused by the geometry
of the Fermi-surface of the cluster, within the IBC \cite{note_1}.
For finite $N_\alpha$ there will be in general an intermediate
temperature range where the specific heat is roughly
linear, like it is the case for $N_\alpha=10$ in
Fig.\ \ref{1D_U2}(b) for temperatures
$0.1t\stackrel{<}{\sim}T\stackrel{<}{\sim}0.3t$. This
feature of the specific heat data obtained with the
IBC may be used, e.g., to estimate the inverse effective mass.

In Fig.\ \ref{2D_U2} we present the specific heat per site,
as a function of temperature, for 2D clusters with $N_s=8,10$,
$U=2t$, $n=0.8$ and $N_\alpha=1$ and $N_\alpha=16$. The improvement
obtained with the IBC is even more pronounced than in 1D.
The data for periodic boundary conditions is affected so much by
finite size errors that is does not allow for a reliable estimate
of the specific heat for $T<2t$. For $N_\alpha=16$ the finite
size corrections are, on the other side, practically absent for
$T>T^*(N_s=10,\epsilon\sim1\%)\sim0.72t$.

In conclusion we have shown that the integration over boundary
conditions (IBC) can be used to reduce substantially
the finite size corrections occuring in exact diagonalization studies.
The data obtained by the IBC show typically a smooth behaviour
as a function of cluster size,
even in 2D where the cluster geometry may vary widely
from cluster to cluster. This smooth behaviour often allows for
{\it quantitative} estimates of the finite size corrections of the
data obtained by the IBC, even in 2D. One obtains furthermore certain
qualitative improvements with the IBC, like a linear specific heat
for the Hubbard model on one- and two dimensional clusters.


This work was
supported by the Deutsche Forschungsgemeinschaft, the Graduiertenkolleg
``Festk\"{o}rperspektroskopie'' and by the European Community Human Capital
and Mobility program.

%
%
%
\begin{figure}
\caption{The ground-state energy, $e_0$, per site as a function
         of particle densitiy $n$.
         Plotted are the results for the 2D Hubbard model
         and $U = 4t$. The filled circles and squares denote
         the results for clusters with
         periodic boundary conditions only, $N_\alpha=1$, for
         clusters with $N_s=8$ and $N_s=10$ sites respectively.
         The long-dashed and the continuous line are the
         respective results for $N_\alpha=25$. Also shown
         (stars) is
         data obtained by projection Monte Carlo for a
         $10\times10$ system by
         Furukawa and Imada \protect\cite{F_I}.
\label{2D_U8}
             }
\end{figure}
\begin{figure}
\caption{The specific heat, $c_V$, per site,
         as a function of temperature,
         of a 6-site
         chain with periodic boundary condition, $n=0.8$
         and (a) $U=2t$ and (b) $U=16t$.
         Plotted are the exact results (solid line)
         and the results of the kernel polyomial approximaiton
         with (a) $N_{mom}=10,100,1000$ and
         with (b) $N_{mom}=100,1000,10000$.
\label{1D6}
             }
\end{figure}
\begin{figure}
\caption{The specific heat, $c_V$, per site
         as a function of temperature,
         for 1D chains with $N_s=4,6,8,10$,
         $U=2t$, $n=0.8$ and (a) $N_\alpha=1$
         and (b) $N_\alpha=10$.
         For $N_s=4,6$ the data are exact,
         for $N_s=8,10$ the KPA has been used
         with $N_{mom}=1000$ and
         $N_r=100,10$ for $N_s=8,10$
         respectively.
\label{1D_U2}
             }
\end{figure}
\begin{figure}
\caption{The specific heat, $c_V$, per site
         as a function of temperature,
         for 2D clusters with $N_s=8,10$,
         $U=2t$, $n=0.8$. Plotted are the
         data for $N_\alpha=1$ (dotted/dashed line)
         and $N_\alpha=16$ (dashed-dotted/solid line).
         The KPA has been used
         with $N_{mom}=1000$ and
         $N_r=10,8$ for $N_\alpha=1,16$
         respectively.
\label{2D_U2}
             }
\end{figure}

\begin{references}
\bibitem{note_1} With this we mean that for $U=0$ the Fermi surface
                 of the thermodynamic limit is {\it exactly} reproduced
                 by the IBC, for any finite cluster. At finite $U>0$ there
                 may be subtle deformations of the Fermi-surface in the
                 thermodynamic limit due to the
                 interaction, which are, of course, not exactly reproduced
                 by the IBC.
\bibitem{note_2} For 2D clusters $N_\alpha=25=5^2$ corresponds
                 to a grid of 5 different boundary conditions
                 both in $\vec L_1$ direction and in
                 $\vec L_2$ direction. Here $\vec L_{1,2}$
                 are the vectors of periodicity of the cluster.
                 For $N_s=8$ one has $\vec L_1=(2,2)$ and
                 $\vec L_2=(-2,2)$ and for $N_s=10$ one has
                 $\vec L_1=(3,1)$ and $\vec L_2=(-1,3)$.
\end{references}
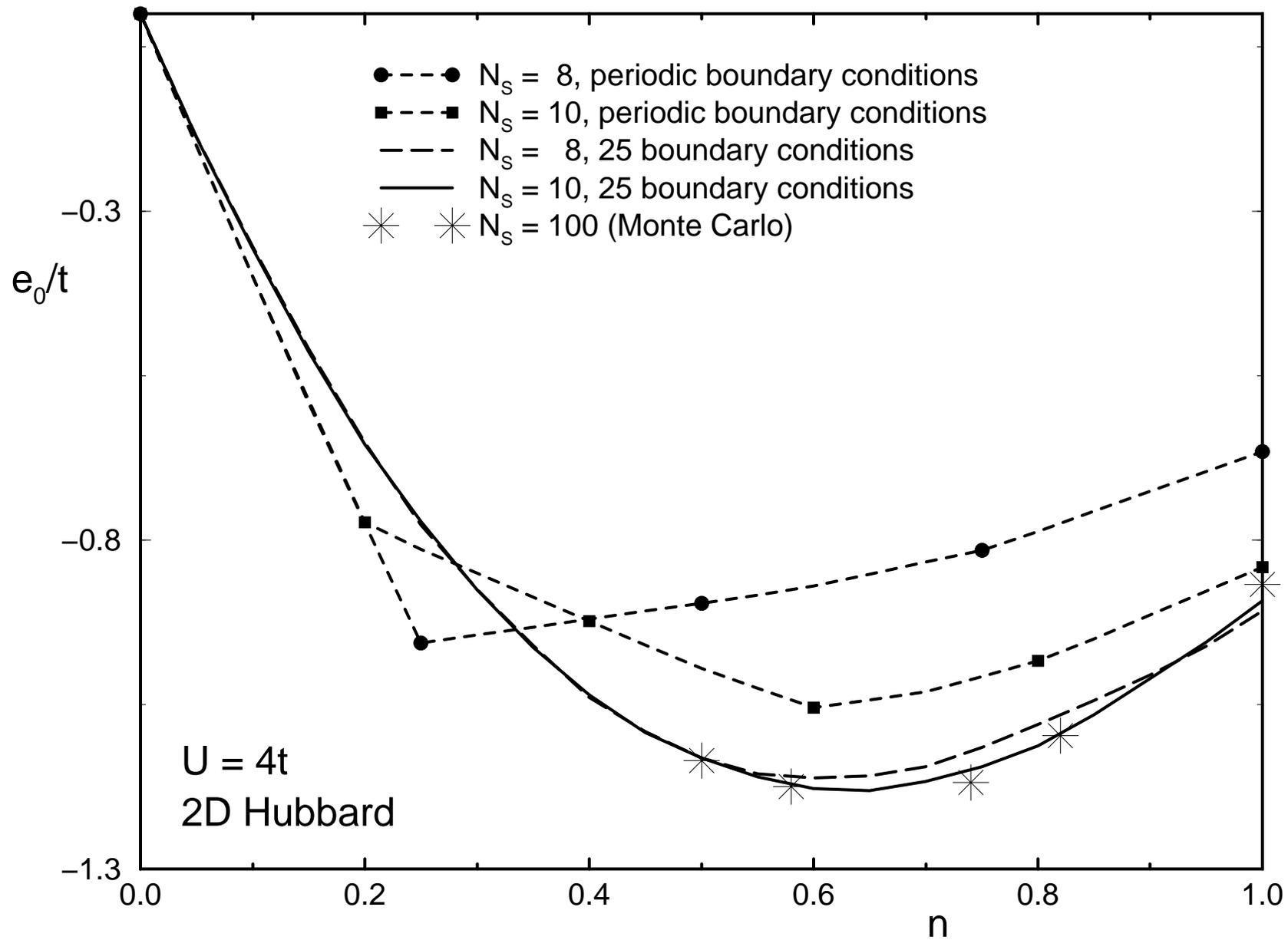
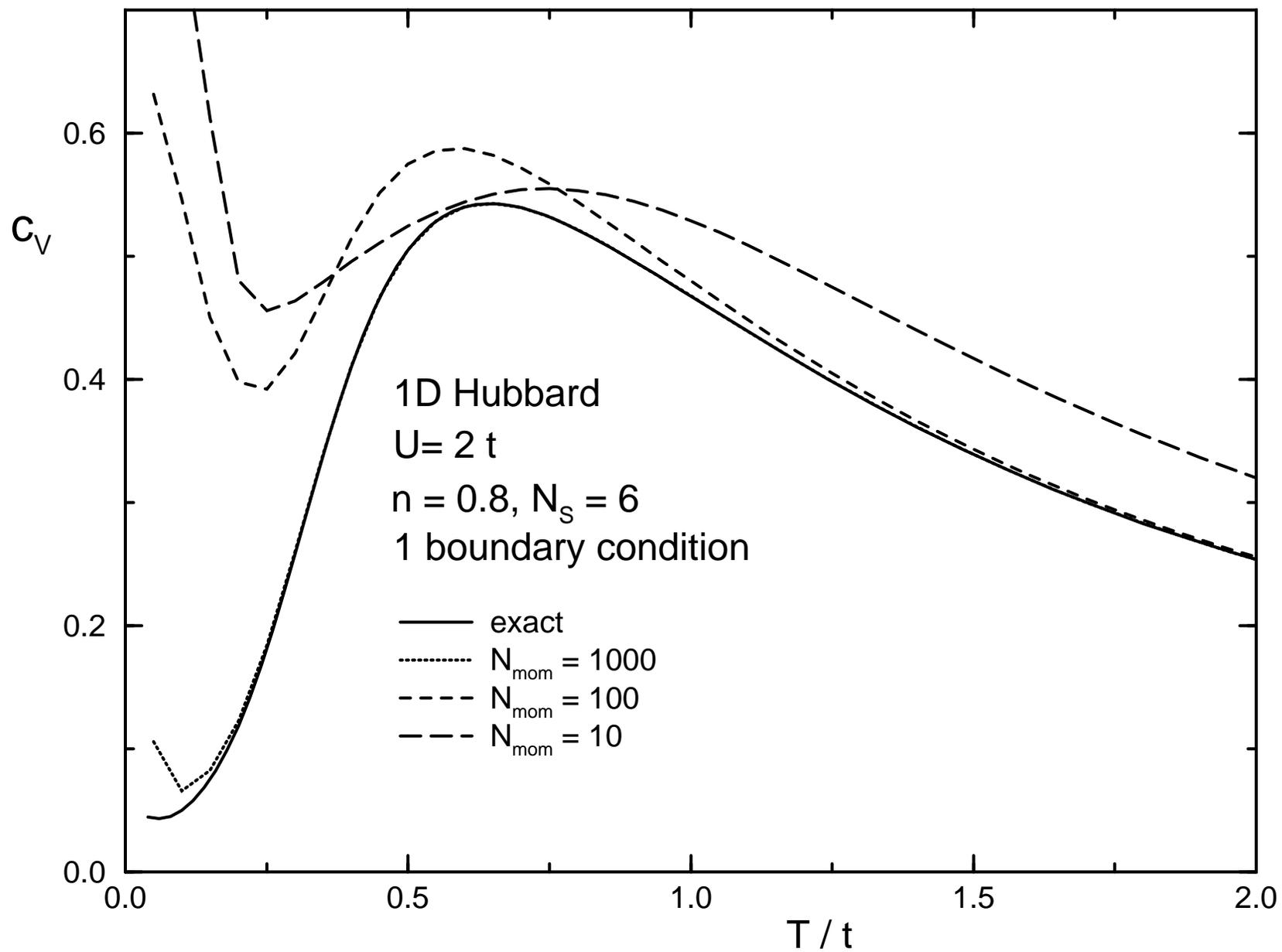
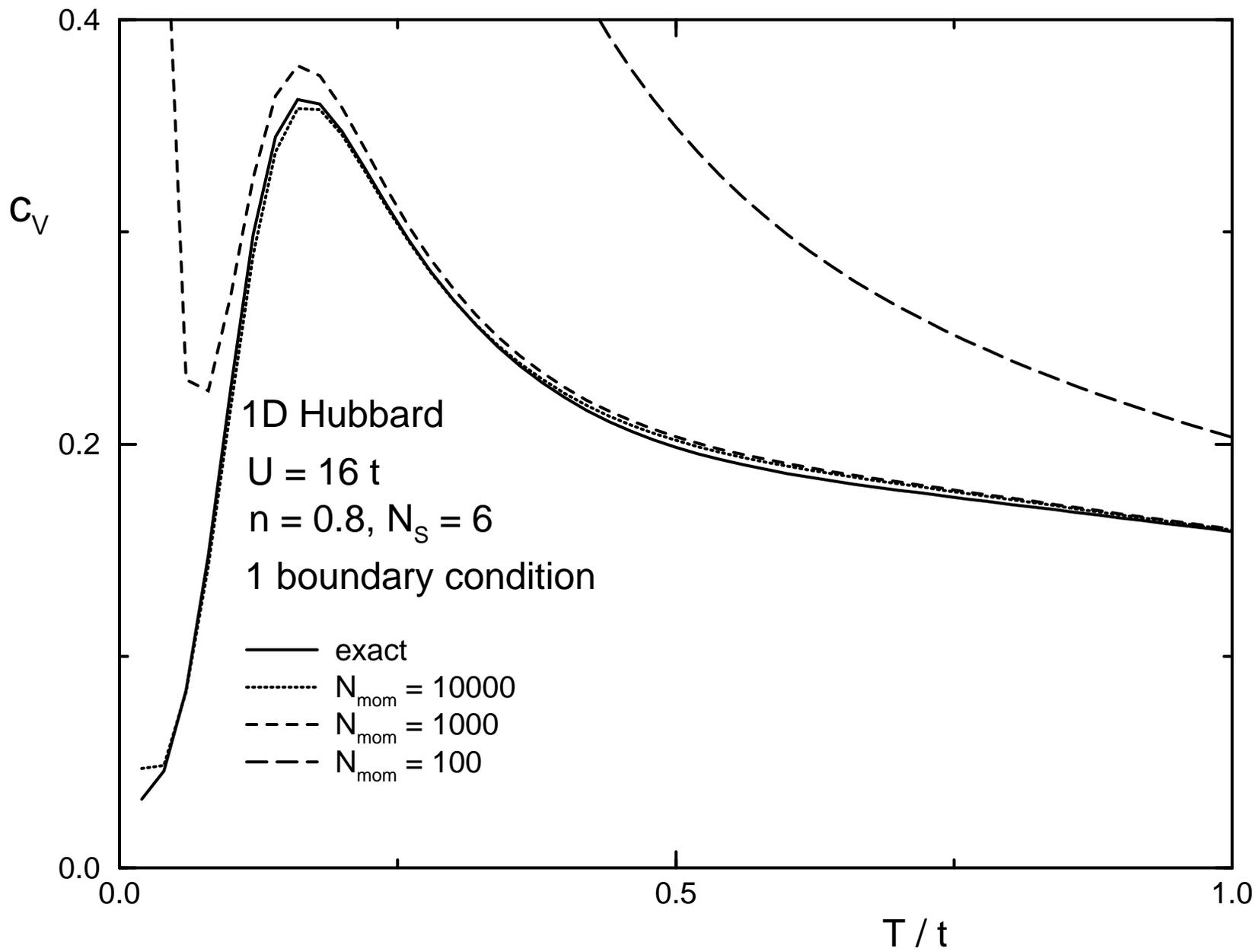
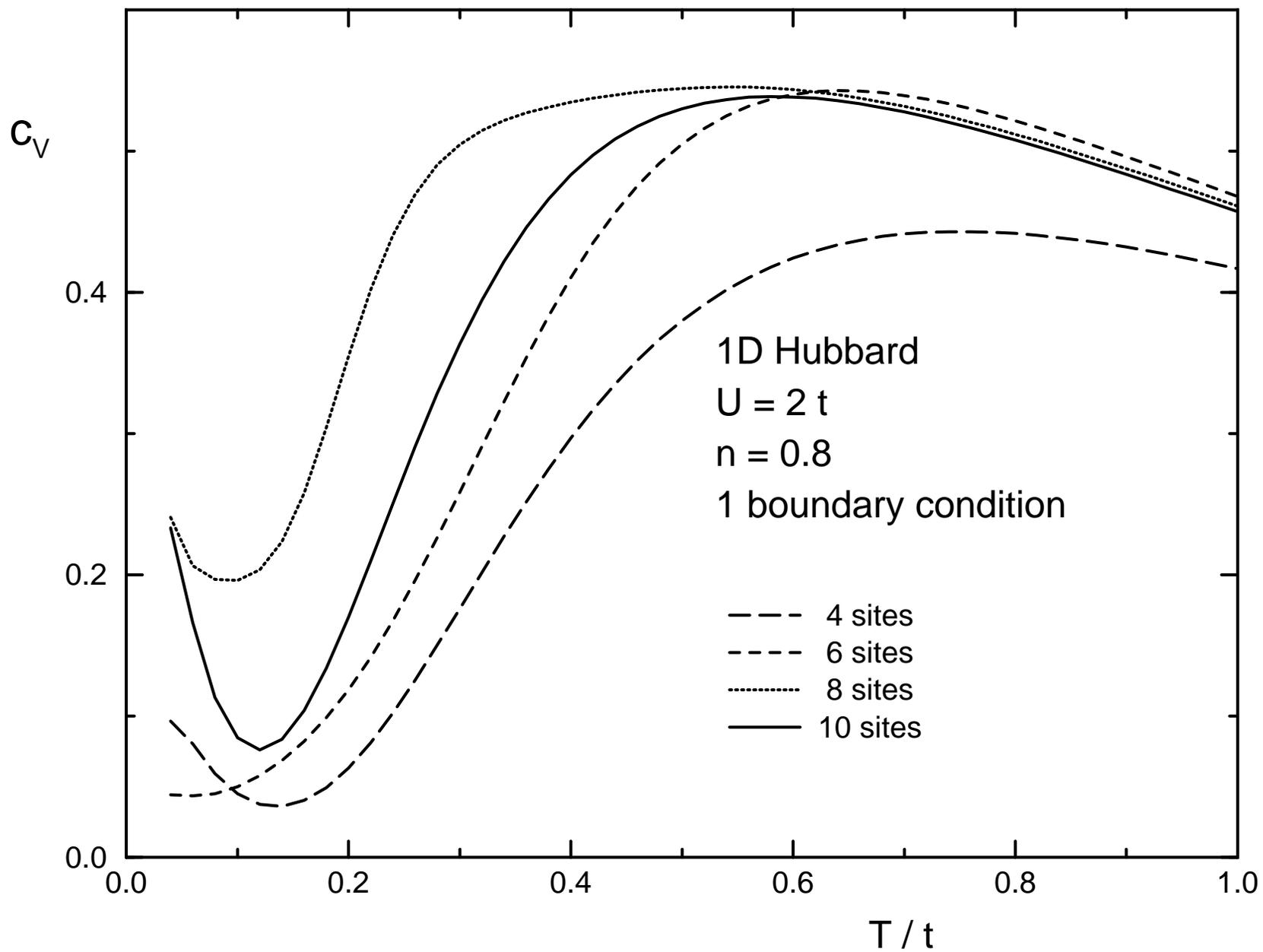
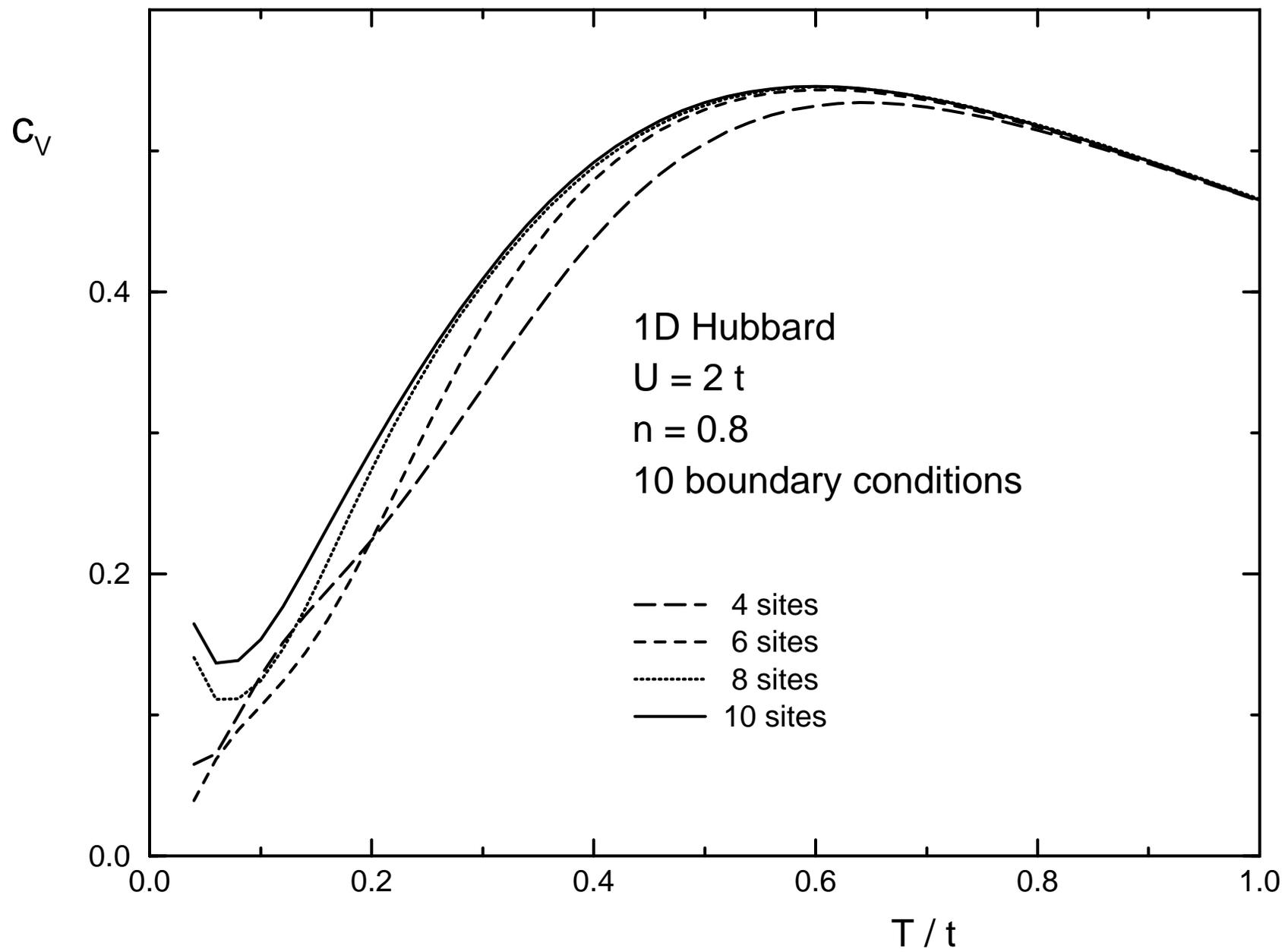
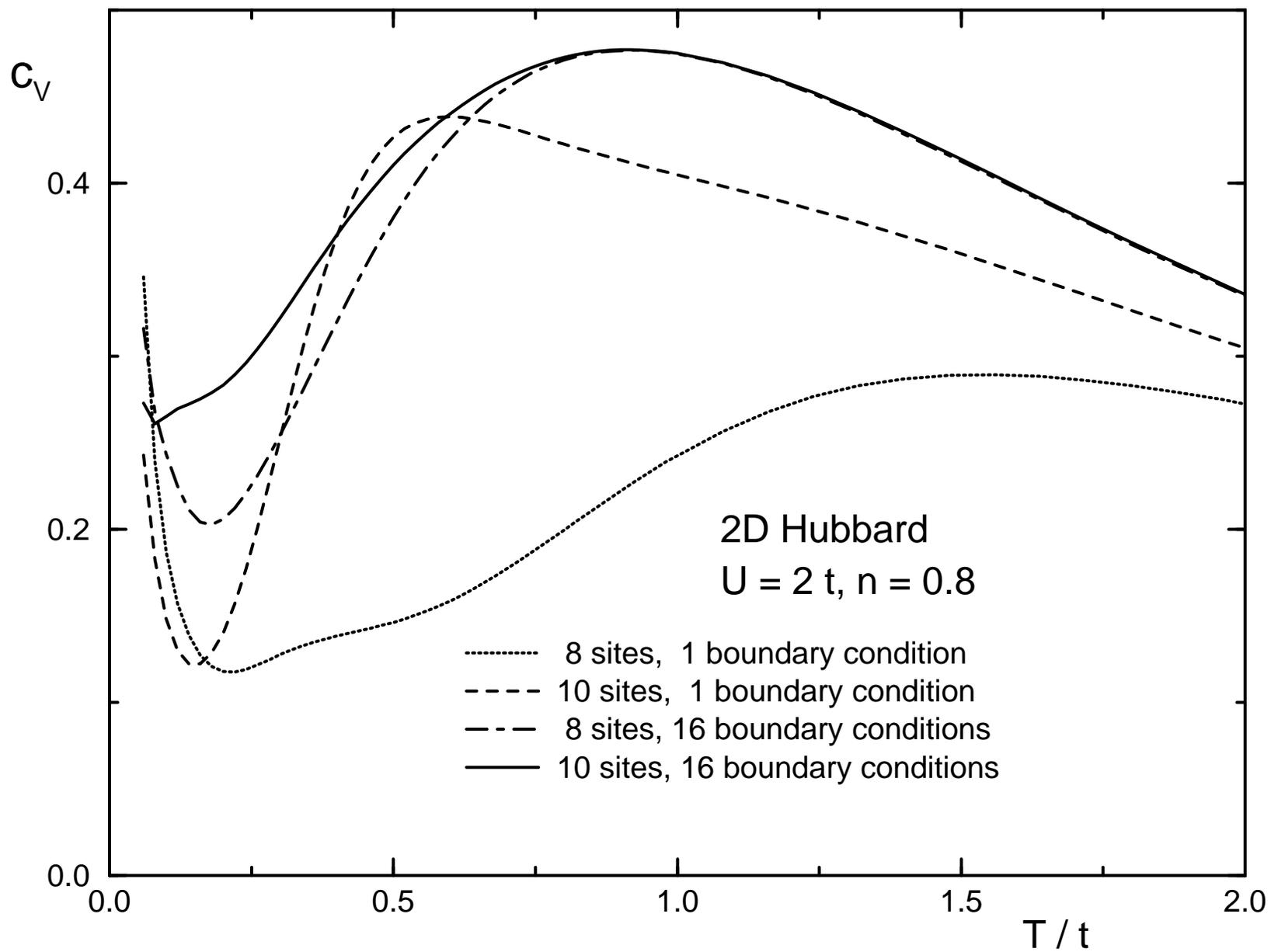
\end{document}